\documentstyle[prl,aps,epsf,floats]{revtex}

\begin{document} 
\draft
\twocolumn[
\hsize\textwidth\columnwidth\hsize\csname@twocolumnfalse\endcsname

\title{Sign reversals of the quantum Hall effect and helicoidal
  magnetic-field-induced spin-density waves in quasi-one-dimensional
  organic conductors}

\author{ N. Dupuis\cite{leave} and Victor M. Yakovenko}
\address{Department of Physics and Center for Superconductivity
  Research, University of Maryland, College Park, MD 20742-4111 }
\date{{\bf cond-mat/9712216}, December 17, 1997}  
\maketitle

\begin{abstract}
  We study the effect of umklapp  scattering on the
  magnetic-field-induced spin-density-wave phases, which are
  experimentally observed in the quasi-one-dimensional organic
  conductors of the Bechgaard salts family. Within the framework of
  the quantized nesting model, we show that umklapp  processes may
  naturally explain sign reversals of the quantum Hall effect (QHE)
  observed in these conductors. Moreover, umklapp  scattering can
  change the polarization of the spin-density wave (SDW) from linear
  (sinusoidal SDW) to circular (helicoidal SDW). The QHE vanishes in the
  helicoidal phases, but a magnetoelectric effect appears.  These
  two characteristic properties may be utilized to detect the  
  magnetic-field-induced helicoidal SDW phases experimentally.
\end{abstract}
]

The organic conductors of the Bechgaard salts family (TMTSF)$_2$X (where
TMTSF stands for tetramethyltetraselenafulvalene) exhibit a rich phase
diagram when temperature, magnetic field, or pressure are varied.  In
three members of this family (X=ClO$_4$, PF$_6$, ReO$_4$), a moderate
magnetic field above several Tesla destroys the metallic phase and induces
a series of spin-density-wave (SDW) phases separated by first-order
phase transitions \cite{rev}.  Because of a strong quasi-one-dimensional
anisotropy (the typical ratio of the electron transfer integrals in
the three crystal directions is $t_a:t_b:t_c=3000:300:10$ K), the
Fermi surfaces of these materials are open.  According to the
so-called quantized nesting model (QNM)
\cite{rev,Heritier84}, the formation of the
magnetic-field-induced spin-density waves (FISDW) results from an 
interplay between the nesting
properties of the Fermi surface and the quantization of the electronic
orbits in magnetic field. The wave vector of
a FISDW adjusts itself to the magnetic field so that unpaired electrons
completely fill an integer number of Landau levels, thus the Hall
effect is quantized \cite{Poilblanc87,Yakovenko91}. The standard QNM
\cite{Heritier84} predicts the Hall plateaus of the same
sign, referred to as positive by convention, which agrees with most
experiments. However, at certain pressures, a negative Hall effect
is also observed \cite{Ribault85,Cooper89,Balicas95}.  In order to
explain the sign reversals of the QHE, Zanchi and
Montambaux \cite{Zanchi96} invoke the variation of the electron
dispersion law with pressure.

In this Letter, we study the effects of umklapp  scattering on the
FISDW phases within the framework of the QNM.  Because the electron
band in the (TMTSF)$_2$X materials is half-filled, the electrons are
allowed to transfer the momentum $4k_F$ along the chains ($k_F$ being
the Fermi momentum) to the crystal lattice.  Therefore, the interaction
between electrons should include not only forward ($g_2$) and backward
($g_1$) scattering amplitudes, but also umklapp  scattering amplitude
($g_3$) \cite{Barisic81}.  We demonstrate that, in the presence of
umklapp  interaction, FISDW phases with a negative QHE appear.
This effect provides an alternative explanation for the sign reversals
of the QHE observed in the Bechgaard salts.  It differs from the one
suggested by Zanchi and Montambaux \cite{Zanchi96} in invoking the
pressure dependence of $g_3$ rather than the electron band structure.
The umklapp  scattering amplitude $g_3$ is sensitive to pressure,
because it is related to the dimerization in the crystal structure of
the TMTSF chains.  Moreover, we show that the polarization of the FISDW
may change from linear (sinusoidal SDW) to circular (helicoidal SDW)
because of umklapp  interaction. The QHE vanishes in the helicoidal phases,
but a magnetoelectric effect appears. These two properties are
characteristic of the helicoidal phases and can be utilized to detect them
experimentally. The effect
of umklapp  on the FISDW phases was studied before by Lebed'
\cite{Lebed90,Lebed91a} using rather crude approximations, but the
helicoidal phases and the sign reversals of the QHE were not
discussed.

In the vicinity of the Fermi energy, the electron dispersion law in
the Bechgaard salts is approximated as
\begin{equation}
  E(k_x,k_y) = v_F(|k_x| -k_F) + t_\perp(k_yb),
\label{disp}
\end{equation}
where $k_x$ and $k_y$ are the electron momenta along and across the
one-dimensional chains of TMTSF, and $\hbar=1$. In Eq.\ (\ref{disp}),
the longitudinal electron dispersion is linearized in $k_x$ in the
vicinity of the two one-dimensional Fermi points $\pm k_F$, and
$v_F=2at_a\sin(k_Fa)$ is the corresponding Fermi velocity, $a$ being
the lattice spacing along the chains.  For the transverse electron
dispersion, a tight-binding approximation is used:
\begin{eqnarray}
t_\perp(k_yb) &=& -2t_b\cos(k_yb) -2t_{2b}\cos(2k_yb) \nonumber \\  
&& -2t_{3b}\cos(3k_yb)-2t_{4b}\cos(4k_yb),
\end{eqnarray}
where $b$ is the distance between the chains. The electron dispersion in
the third direction along the $z$ axis is not important for the
following and is not considered here.

When a magnetic field $H$ is applied along the $z$ axis perpendicular
to the $(x,y)$ plane, it quantizes the transverse electron motion into
the Wannier-Stark ladder \cite{ND95}. Consequently, the static spin
susceptibility $\chi_0({\bf q})$, calculated at a wave vector ${\bf
  q}=(q_x,q_y)$ in the absence of interaction between electrons,
diverges logarithmically at quantized values of the longitudinal
wave vector $q_x^{(n)}=2k_F+nG$ ($n$ integer) \cite{rev,Montambaux85}:
\begin{equation}
\chi_0({\bf q})= \sum_n I_n^2(q_y)\, \chi_{1D}(q_x-nG). 
\label{chi0}
\end{equation}
Here $G=eHb/\hbar $ is the characteristic wave vector
of the magnetic field ($e$ is the electron charge), and
$\chi_{1D}(q_x)$ is the susceptibility of a 1D system without
interaction. The coefficients $I_n$ depend on the transverse
dispersion law of electrons:
\begin{equation}
I_n(q_y)= \langle  e^{inu + \frac{i}{v_FG}
\lbrack T_\perp(u+q_yb/2) + T_\perp(u-q_yb/2) \rbrack } \rangle,
\end{equation}
where $T_\perp(u)=\int _0^u du' t_\perp(u')$, and $\langle\cdots\rangle$
denotes averaging over $u$.  In the absence of umklapp  scattering, the
transition temperature of the FISDW is determined by the 
Stoner criterion $1-g_2\chi_0(Q_x^{(N)},Q_y)=0$. The quantized
longitudinal wave vector $Q_x^{(N)}=2k_F+NG$ and the
transverse wave vector $Q_y$ are selected to maximize the
transition temperature $T_c^{(N)}$ at a given magnetic field. Except when
$N=0$, $Q_y$ is incommensurate: $Q_y\ne \pi /b$. The
integer parameter $N$ also determines the quantum Hall conductivity in
the FISDW phase: $\sigma_{xy}=-2Ne^2/h$ per one layer of the TMTSF
molecules \cite{Poilblanc87,Yakovenko91}.  As the magnetic field
increases, the value of $N$ changes, which leads to a cascade of
FISDW transitions \cite{rev,Heritier84}.

Umklapp  scattering mixes the wave vectors $Q_x^{(N)}$ and
$Q_x^{(N)}-4k_F=-Q_x^{(-N)}$, thus two SDWs, with the wave vectors
${\bf Q}_N=(Q_x^{(N)},Q_y)$ and ${\bf Q}_{-N}=(Q_x^{(-N)},-Q_y)$, form
simultaneously \cite{Lebed90,Lebed91a}. In the random-phase
approximation, the critical temperature $T_c^{(N)}$ is determined by
the modified Stoner criterion \cite{Lebed-a}:
\begin{eqnarray}
[1-g_2\chi_0({\bf Q}_N)]
[1-g_2\chi_0({\bf Q}_{-N})] && \nonumber \\  
 -g_3^2 \chi_0({\bf Q}_N) \chi_0({\bf Q}_{-N}) &=&  0.
\label{stoner}
\end{eqnarray}
Of the two integers $N$ and $-N$, we select $N$ such that $\chi_0({\bf
  Q}_N)>\chi_0({\bf Q}_{-N})$ to label each FISDW phase.

Below $T_c^{(N)}$, the system is characterized by the order parameters
$\Delta _{\beta N,\alpha}$:
\begin{equation}
\langle \hat{\psi}^\dagger _{\alpha,\uparrow}({\bf r})
\hat{\psi}_{-\alpha,\downarrow}({\bf r}) \rangle
=\sum_{\beta=\pm}\Delta _{\beta N,\alpha}
e^{-i\alpha{\bf r}\cdot{\bf Q}_{\beta N}},
\label{eq:psi}  
\end{equation}
where ${\bf r}=(x,y)$ is the spatial coordinate. In Eq.\ 
(\ref{eq:psi}), the index $\beta=\pm$ labels the wave vectors ${\bf
  Q}_{\pm N}$.  The operators $\psi ^{(\dagger )}_{\alpha ,\sigma }$
annihilate (create) electrons with spin $\sigma $ and momenta close to
$\alpha k_F$ ($\alpha =\pm $). The electron
spin density has a nonzero expectation value varying in space
\cite{mPsi}:
\begin{eqnarray}
\langle S_x({\bf r})\rangle &=&
  \sum_{\beta=\pm} m^{(x)}_{\beta N} 
  \cos( {\bf r}\cdot{\bf Q}_{\beta N} + \theta_{\beta N}^{(x)} ),
\nonumber \\ 
\langle S_y({\bf r})\rangle &=& 
  \sum_{\beta=\pm} m^{(y)}_{\beta N} 
 \cos( {\bf r}\cdot{\bf Q}_{\beta N} + \theta_{\beta N}^{(y)} ).
\label{SxSy}
\end{eqnarray}
When $\theta _{\beta N}^{(x)}=\theta _{\beta N}^{(y)}$, Eq.\ 
(\ref{SxSy}) describes sinusoidal SDWs. When $m^{(x)}_{\beta
  N}=m^{(y)}_{\beta N}$ and $\theta_{\beta N}^{(x)}=\theta_{\beta
  N}^{(y)}\pm\pi/2$, it describes helicoidal SDWs.
The spin polarization vector $\langle{\bf S}({\bf r})\rangle$ of a
generic helicoidal SDW rotates in the plane perpendicular to a vector
{\bf n} when the coordinate {\bf r} varies. In our case, because of
the Zeeman effect, the vector {\bf n} aligns itself with the magnetic
field {\bf H}, thus the spin polarization (\ref{SxSy}) rotates in the
$(x,y)$ plane.

The actual polarization of the SDWs is determined by minimizing the
free energy of the system.  In terms of the linear combinations of the
order parameters:
\begin{equation}
  \tilde \Delta_{\beta N,\alpha} = I_{\beta N}(Q_y)\,
  (g_2\Delta _{\beta N,\alpha} + g_3\Delta _{-\beta N,-\alpha}),
\label{eq:Delta}
\end{equation}
the Landau expansion of the free energy in the vicinity of $T_c^{(N)}$
for the phase $N$ has the following form \cite{Lebed-b}:
\begin{eqnarray}
F_N &=& \sum_{\alpha} \Bigl \lbrack \,
\sum_\beta  A_{\beta N} |\tilde \Delta_{\beta N,\alpha}|^2
+ B (\tilde \Delta_{N,\alpha}\tilde \Delta ^*_{-N,-\alpha}+{\rm c.c.} ) 
\nonumber\\ && + (K/2) \sum_\beta |\tilde \Delta _{\beta N,\alpha }|^4 
+ 2K |\tilde \Delta_{N,\alpha}\tilde \Delta_{-N,\alpha}|^2 \Bigr \rbrack,
\label{FN}
\end{eqnarray}
where the coefficients are
\begin{eqnarray}
A_{\beta N} &=& \frac{1}{I_{\beta N}^2(Q_y)}
\left( \frac{g_2}{g_2^2-g_3^2} - \chi_0({\bf Q}_{\beta N}) \right),
\label{A} \\ 
B &=& -g_3 /  I_{N}(Q_y)I_{-N}(Q_y)(g_2^2-g_3^2),
\label{B} \\
K &=& 7\zeta(3) / 16 \pi^3 v_F b T^2, \quad \zeta(3)\simeq1.20.
\label{K}
\end{eqnarray}

As long as the quadratic part of the free energy $F_N$ (\ref{FN}) is
positively defined, the metallic state is stable. The second-order
phase transition into a FISDW state takes place when the determinant
of the quadratic part of Eq.\ (\ref{FN}) vanishes: $A_NA_{-N}=B^2$.
With the coefficients $A$ and $B$ given by Eqs.\ (\ref{A}) and
(\ref{B}), this condition is equivalent to the Stoner equation
(\ref{stoner}).  Minimizing the free energy $F_N$ (\ref{FN}) with
respect to $\tilde \Delta_{\pm N,\pm}$ at $T<T_c^{(N)}$, we find two types of
solutions depending on the value of $g_3$. For small $g_3$, when
$\sqrt{2}|B|<|A_N-A_{-N}|$, we find a solution where
$|\tilde \Delta_{N,+}|=|\tilde \Delta_{N,-}|$ and $|\tilde
\Delta_{-N,+}|=|\tilde \Delta_{-N,-}|$,
which corresponds to two sinusoidal SDWs.  When $g_3$
exceeds a certain critical value so that $\sqrt{2}|B|>|A_N-A_{-N}|$,
the minimum of the free energy is reached at $\tilde \Delta_{N,-}$=$\tilde
\Delta_{-N,+}=0$ and $|\tilde \Delta_{N,+}|,|\tilde \Delta_{-N,-}|\neq 0$, 
which corresponds to two
helicoidal SDWs of opposite chiralities. Using the method of Ref.\ 
\cite{Yakovenko91}, we find that the QHE is quantized in the
sinusoidal phase: $\sigma_{xy}=-2Ne^2/h$, but vanishes in the
helicoidal phase. We also find that, when the magnetic field is
varied, phase transitions between adjacent FISDW phases, whether
sinusoidal or helicoidal, are of the first order \cite{ND}.  The
conclusion of Lebed' \cite{Lebed90} that, in the presence of umklapp 
scattering, adjacent FISDW phases are separated by two second-order phase
transitions and an intermediate phase with coexistence of four SDWs is
incorrect, because he did not consider helicoidal SDWs.

Having gained an analytical insight into the problem, we study the phase
diagram in the presence of umklapp  scattering numerically. It is convenient
to characterize the interaction by dimensionless coupling constants
$\tilde g_2=g_2/\pi v_Fb$ and $\tilde g_3=g_3/\pi v_Fb$. We vary the
ratio $\tilde g_3/\tilde g_2$ while keeping the sum constant: $\tilde
g_2+\tilde g_3=2/\ln(2\gamma E_0/\pi T_c^{(\infty)})$, where
$T_c^{(\infty)}=12$ K is the transition temperature for an infinite
magnetic field, $E_0$ is an ultraviolet cutoff of the order of $t_a$,
and $\gamma \simeq 1.781$ is the exponential of the Euler constant.
The calculations are performed for $t_b=300$ K, $t_{2b}=20$ K,
$t_{3b}=0$ K, $t_{4b}=0.75$ K, and $E_0=2000 $ K.  The transition
temperature $T_c^{(N)}$ is obtained numerically from Eq.\ 
(\ref{stoner}), and the polarizations of the SDWs for $T<T_c^{(N)}$ are 
determined by minimizing the free energy $F_N$ given by Eq.\ 
(\ref{FN}).

A very small $g_3$ does not change the phase diagram qualitatively
compared to the case $g_3=0$. Now the main SDW at the wave vector
${\bf Q}_N$ coexists with a weak SDW at the wave vector ${\bf
  Q}_{-N}$. In general, the value of $Q_y$ that maximizes $\chi_0({\bf
  Q}_N)$ does not maximize $\chi_0({\bf Q}_{-N})$, so $\chi_0({\bf
  Q}_{-N})\ll\chi_0({\bf Q}_N)$. As a result, the SDW amplitude at the
wave vector ${\bf Q}_{-N}$ is very small, and the polarizations of the
SDWs are linear. The values of $N$ follow the usual ``positive''
sequence $N=\ldots,6,5,4,3,2,1,0$ with increasing magnetic field.

\begin{figure}
\epsfxsize 7.3cm 
\epsffile[170 550 395 700]{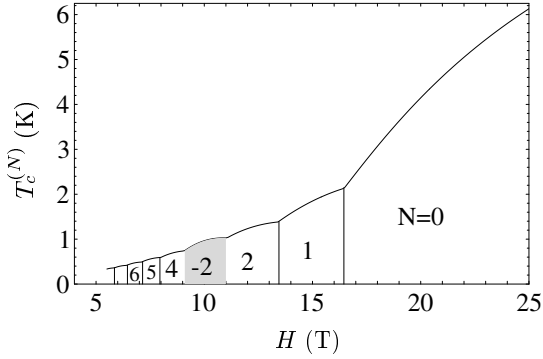}
\caption{ Phase diagram for $g_3/g_2=0.03$ ($\tilde g_2\simeq 0.37$
  and $\tilde g_3\simeq 0.01$). The phase $N=3$ is suppressed, and the
  negative commensurate phase with $N=-2$ and $Q_y=\pi/b$ appears in
  the cascade (the shaded area). All the phases are sinusoidal, and
  the Hall effect is quantized: $\sigma_{xy}=-2Ne^2/h$. The vertical
  lines are only guides for the eyes and do not necessarily correspond to
  the actual first-order transition lines. }
\label{Fig:r0v03}
\end{figure}

A larger value of $g_3$ increases the coupling between the two SDWs. This
leads to a strong decrease of the critical temperature or even the 
disappearance of the SDWs. However, for even $N$, there exists a
critical value of $g_3$ above which the system prefers to choose the
transversely commensurate wave vector $Q_y=\pi/b$ for both SDWs. The
reason is that, for even $N$ (as opposed to odd $N$), $Q_y=\pi/b$
corresponds to a local maximum of the susceptibilities {\it and}
$\chi_0(Q_x^{(N)},\pi/b)\simeq \chi_0(Q_x^{(-N)},\pi/b)$. The latter
relation implies that the two SDWs have {\it comparable} amplitudes.
The two susceptibilities are strictly equal at $t_{4b}=0$, but when
$t_{4b}>0$, $\chi_0(Q_x^{(-N)},\pi/b)>\chi _0(Q_x^{(N)},\pi/b)$ (this result
holds also for $t_{3b}\neq0$) \cite{Zanchi96}. This yields a negative
Hall plateau, provided the SDWs are sinusoidal. Thus, for
$g_3/g_2=0.03$ ($\tilde g_2\simeq 0.37$ and $\tilde g_3\simeq 0.01$),
we find the sequence $N=\ldots,6,5,4,-2,2,1,0$ (see Fig.\ 
\ref{Fig:r0v03}).  The phase $N=3$ is suppressed, and the negative
commensurate phase with $N=-2$ and $Q_y=\pi/b$ appears in the cascade.
All the phases are sinusoidal, so the Hall effect is quantized
($\sigma_{xy}=-2Ne^2/h$).

The strength of umklapp  scattering is very sensitive to pressure,
because hydrostatic pressure reduces the dimerization in the crystal
structure of the TMTSF chains, which diminishes $g_3$. Therefore, we
conclude that the sign reversals of the QHE can be induced by varying
pressure. In our simplified model, this effect requires $t_{4b}>0$.
Our results provide a new explanation of the negative Hall plateaus in
the Bechgaard salts.

\begin{figure}
\epsfxsize 7.3cm 
\epsffile[170 550 395 700]{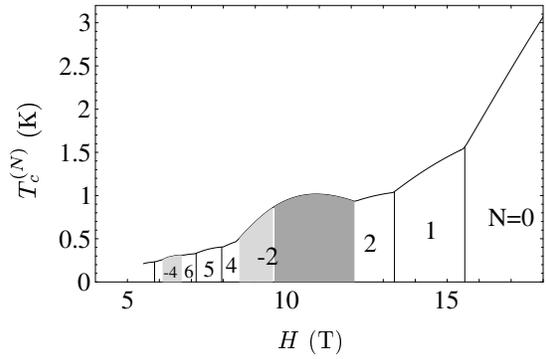}
\caption{ Phase diagram for $g_3/g_2=0.06$ ($\tilde g_2\simeq 0.36$
  and $\tilde g_3\simeq 0.02$). Two negative phases, $N=-2$ and
  $N=-4$, are observed (the shaded areas). The phase $N=-2$ splits
  into two subphases: helicoidal (the dark shaded area) and sinusoidal
  (the light shaded area). }
\label{Fig:r0v06}
\end{figure}

If $g_3/g_2$ is increased to 0.06, a second negative phase ($N=-4$)
appears, and the cascade becomes $N=\ldots,8,7,-4,6,5,4,-2,2,1,0$ (see
Fig.\ \ref{Fig:r0v06}). As discussed in Ref.\ \cite{Zanchi96}, $N=-2$ and
$N=-4$ correspond to the two negative QHE phases observed in experiments
\cite{Cooper89,Balicas95}. As shown in Fig.\ \ref{Fig:r0v06}, the phase
$N=-2$ splits into two subphases: helicoidal and sinusoidal. {\it Not
  only does umklapp  scattering stabilize negative phases, but, as
  $g_3$ increases, these negative phases are likely to become
  helicoidal.}  Therefore, in order to observe the helicoidal phase
experimentally, it would be desirable to stabilize the negative phase
$N=-2$ at the lowest possible pressure (which corresponds to the
strongest $g_3$). In (TMTSF)$_2$PF$_6$, the pressure has to be higher than
6 kbar, since the FISDW cascade disappears below this pressure
\cite{rev}. In the experiment \cite{Balicas95}, where the phase $N=-2$
has been observed at 8.3 kbar, the pressure could be reduced by only
about 2 kbar.  Nevertheless, such a pressure reduction could induce a
significant increase of $g_3$.  (TMTSF)$_2$ReO$_4$, where the sign
reversals of the QHE have been observed under pressure \cite{Kang91},
could also be a good candidate for detecting helicoidal phases.

\begin{figure}
\epsfysize 4.4cm 
\epsffile[90 375 420 580]{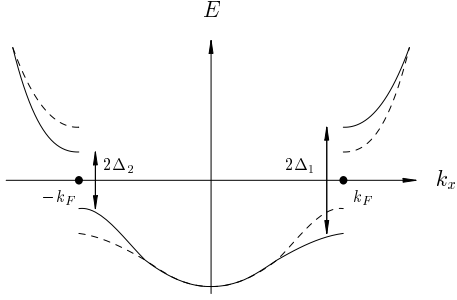}
\caption{Spectrum of electronic excitations in the helicoidal FISDW phase. 
  The solid (dashed) line corresponds to up (down) spins. }
\label{Fig:spec}
\end{figure}

The helicoidal FISDW phases exhibit a kinetic magnetoelectric effect and
vanishing QHE. The magnetoelectric effect may exist if time-reversal
and space-inversion symmetries are broken \cite{Landau}. 
Gor'kov and Sokol found the kinetic magnetoelectric effect
for a single helicoidal SDW \cite{Gorkov87}. The effect also
exists in the presence of two helicoidal SDWs of opposite chiralities,
provided their amplitudes are not equal.  An electric current along
the chains, $j_x$, induces a uniform magnetization {\bf M} along the
vector {\bf n} that characterizes the spin polarization of a
helicoidal SDW. In our case, the vector {\bf n} is parallel to the
magnetic field {\bf H}, which is oriented along the $z$ axis, thus
$M_z\propto j_x$.  (Here $M_z$ is the additional spin magnetization
induced by $j_x$ in excess of the magnetization induced by the
magnetic field without $j_x$.)  The effect can be understood by
considering the spectrum of electronic excitations in the helicoidal
FISDW phase shown in Fig.\ \ref{Fig:spec}.  The $+k_F$ electrons with
spin up and the $-k_F$ electrons with spin down have the energy gap
$\Delta_1=|\tilde \Delta_{N,+}|$, whereas the $+k_F$ electrons with spin down
and the $-k_F$ electrons with spin up have the different energy gap
$\Delta_2=|\tilde \Delta_{-N,-}|$. To produce a current $j_x$ along the
chains, electrons need to be 
transferred from $-k_F$ to $+k_F$. For $\Delta_1\neq\Delta_2$
($\Delta_1\neq\Delta_2$ if $t_{4b}\neq 0$), this
redistribution of electrons results in a uniform magnetization $M_z$:
\begin{eqnarray}
j_x &=& ev_F \sum_\sigma (\delta n_{+,\sigma}-\delta n_{-,\sigma}),
\nonumber \\ 
M_z &=& \frac{g\mu_B}{2} \sum_{\alpha=\pm}
(\delta n_{\alpha,\uparrow} - \delta n_{\alpha,\downarrow}),
\end{eqnarray}
where $\delta n_{\alpha,\sigma}$ is the deviation of the distribution 
function of electrons with spin $\sigma$ and momenta near
$\alpha k_F$ from the equilibrium one, $g$ is the electron
gyromagnetic factor, and $\mu_B$ is the Bohr magneton.  At low
temperature, we can consider solely the electrons excited above the lowest gap
($\Delta_2$ in Fig.\ \ref{Fig:spec}). This implies that $\delta
n_{+,\uparrow }\simeq \delta n_{-,\downarrow }\simeq 0$ and
$M_z/j_x\simeq -g\mu _B/2ev_F$.

In conclusion, we have shown that the negative phases (i.e., with a sign
reversal of the QHE) observed in the Bechgaard salts can be explained by
considering umklapp processes. These phases are characterized by
the coexistence of two linearly polarized SDWs (with the wave vectors
${\bf Q}_N$ and ${\bf Q}_{-N}$) with comparable amplitudes.  We have
also shown that these negative phases are likely to become helicoidal
under low pressure.  The helicoidal phases have no QHE and exhibit a
magnetoelectric effect. The latter effect can be utilized to detect the
helicoidal phases experimentally by looking for a spin magnetization
proportional to the current along the chains.

This work was partially supported by the NSF under Grant DMR--9417451
and by the Packard Foundation.

\vspace{-1.5\baselineskip}


\begin{references}
\bibitem{Lebed-a} A similar equation, with $Q_y=\pi /b$ and an
  approximate expression for $\chi_0$, has been obtained in Refs.\ 
  \cite{Lebed90,Lebed91a}.
\bibitem{mPsi} The parameters of the SDWs in Eq.\ (\ref{SxSy}) are
  related to the order parameters in Eq.\ (\ref{eq:psi}):
  $m^{(x)}_{\beta N} \exp(i\theta_{\beta N}^{(x)}) =\Delta _{\beta N,+}^*
  +\Delta_{\beta N,-}$ and $m^{(y)}_{\beta N} \exp(i\theta_{\beta
    N}^{(y)}) =i(\Delta_{\beta N,+}^* - \Delta_{\beta N,-})$.
\bibitem{Lebed-b} The Landau expansion of the free energy has been
  studied in Ref.\ \cite{Lebed90} for sinusoidal SDWs.
\end{references}
\end{document}